# Crossed Luttinger Liquid Hidden in a Quasi-two-dimensional Material η-Mo$_4$O$_{11}$

X. Du[1], L. Kang[1], Y. Y. Lv[2], J. S. Zhou[1], X. Gu[1], R. Z. Xu[1], Q. Q. Zhang[1], Z. X. Yin[1], W. X. Zhao[1], Y. D. Li[1], S. M. He[3], D. Pei[3], Y. B. Chen[2], M. X. Wang[4,5], Z. K. Liu[4,5], Y. L. Chen[1,3,4,5*], and L. X. Yang[1,6,7*]

[1]*State Key Laboratory of Low Dimensional Quantum Physics, Department of Physics, Tsinghua University, Beijing 100084, China.*
[2]*National Laboratory of Solid State Microstructures Department of Materials Science and Engineering Nanjing University, Nanjing 210093, China.*
[3]*Department of Physics, Clarendon Laboratory, University of Oxford, Parks Road, Oxford OX1 3PU, UK.*
[4]*School of Physical Science and Technology, ShanghaiTech University and CAS-Shanghai Science Research Center, Shanghai 201210, China.*
[5]*ShanghaiTech Laboratory for Topological Physics, Shanghai 200031, China.*
[6]*Frontier Science Center for Quantum Information, Beijing 100084, China.*
[7]*Collaborative Innovation Center of Quantum Matter, Beijing 100084, China*

*e-mail: LXY: lxyang@tsinghua.edu.cn；YLC: yulin.chen@physics.ox.ac.uk;

**Although the concept of Luttinger liquid (LL) that describes a one-dimensional (1D) interacting fermion system[1,2] collapses in higher dimensions, it has been proposed to be closely related to many mysteries including the normal state of cuprate superconductor[3-5], unconventional metal[6,7], and quantum criticality[8,9]. Therefore, the generalization of LL model to higher dimensions has attracted substantial research attention[4,10-12]. Here we systematically investigate the electronic structure of a quasi-2D compound η-Mo$_4$O$_{11}$ using high-resolution angle-resolved photoemission spectroscopy and *ab-initio* calculation. Remarkably, we reveal a prototypical LL behavior originated from the crossing quasi-1D chain arrays hidden in the quasi-2D crystal structure. Our results suggest that η-Mo$_4$O$_{11}$ materializes the long sought-after crossed LL phase[13-15], where the orthogonal orbital components significantly reduce the coupling between intersecting quasi-1D chains and therefore maintain the essential properties of LL. Our finding not only presents a realization of 2D LL, but also provides a new angle to understand non-Fermi liquid behaviors in other 2D and 3D quantum materials.**

Non-Fermi liquids are unconventional metallic systems that cannot be described by Landau's Fermi liquid theory due to the strong electron correlation, which were suggested to be the key to understand some of the mysterious phases in condensed matter physics, such as the normal state of cuprate high-temperature superconductors[16,17], heavy-fermion materials[18,19], and the normal state of itinerant-electron ferromagnets[20]. As an example, due to the reduction of scattering phase-space and electronic screening[1], one-dimensional (1D) electronic systems can harbor Luttinger liquid (LL) state, which exhibits spin-charge separation and power-law scaling of physical properties.

Although the LL theory was developed to describe 1D electron systems, it is believed to be highly relevant to many long-standing mysteries in 2D and 3D quantum materials, such as the normal state of cuprate high-temperature superconductors[3-5], unconventional metals[6,7], and quantum criticality[8,9]. Therefore, it is of fundamental importance to generalize the LL phenomenology to higher dimensions. Many theoretical models suggest that the essential features of LL can be retained in quasi-1D systems consisting of parallel-arrayed 1D chains[21-24] (Fig. 1a), which can be understood by a coupled LL model[11], as exemplified in various experimental demonstrations[25-30]. Such coupled quantum wires are further proposed to host fascinating properties such as non-Abelian quantum Hall states, topological phases, and quantum spin liquid[4,10,12,31].

Particularly, a novel 2D LL phase can be established by weaving 1D atomic chains as proposed in the crossed sliding LL model (Fig. 1b), which is highly relevant to the stripe phase of the cuprates[13,32]. However, the realization of this intuitive theoretical proposal is challenging, as it requires a delicate construction of ordered intercrossing chain arrays. Indeed, except the pioneering effort to pattern carbon nanotubes with only one crossing[15,33], no compelling evidence of a genuinely 2D LL has been identified yet.

Here, combining high-resolution angle-resolved photoemission spectroscopy (ARPES) and *ab-initio* calculation, we systematically investigate the electronic structure of a quasi-2D molybdenum oxide η-$Mo_4O_{11}$. Remarkably, we discover a LL phase that is manifested by a robust power-law scaling behavior of ARPES spectra near the Fermi energy ($E_F$), which is independent of the temperature and momentum direction. By constructing an effective tight-binding model, the low energy electronic structure can be reproduced using three chain arrays featuring orthogonal Mo 4$d$ $t_{2g}$ orbitals, revealing its nature of crossed 1D electronic structure hidden in the quasi-2D crystal structure, in excellent agreement with our experiment. The inter-array coupling in η-$Mo_4O_{11}$ is weak due to the orthogonality between the orbital components of different chain arrays, which naturally materialize the crossed LL model. This system thus not only demonstrate a novel LL phase in a quasi-2D material, but will also provide a new angle to understand non-Fermi liquid behaviors in other 2D and 3D quantum materials.

η-$Mo_4O_{11}$ crystallizes into a monoclinic structure with space group $P2_1/c$. As shown in Fig. 1c, it exhibits a quasi-2D structure consisting of $Mo_6O_{22}$ slabs separated by $MoO_4$ tetrahedrons[34]. The slabs are composed of corner-sharing $MoO_6$ octahedrons, forming chain-like structures along three different directions, as highlighted in Fig. 1d (Supplementary Fig. S1) and manifested by the three natural crystalline directions of the sample (inset of Fig. 1e). Laue diffraction pattern along the (100) plane (Fig. 1e) and X-ray diffraction pattern (Supplementary Fig. S2) confirm the crystal structure and the high quality of our samples[35]. The three-dimensional plot of the experimental band structure in Fig. 1f shows highly dispersive bands crossing $E_F$ and parallel straight lines on the Fermi surface (FS), consistent with previous ARPES measurements[36,37].

Figure 2 shows the detailed electronic structure of η-$Mo_4O_{11}$. The measured FS (Fig. 2a) consists of three groups of straight lines running along different directions (Supplementary Fig. S3). The angle between the C1 and C2 (C2 and C3) FS sheets is 51° (78°), consistent with the natural crystalline directions (inset of Fig. 1e) and implying 1D physics hidden in the quasi-2D crystal structure[38]. The different FS sheets hybridize with each other, inducing gaps at the crossing points, as shown by the zoom-in plots and momentum distribution curves (MDCs) in Fig. 2b. In Fig. 2c, we observe two highly dispersive bands along $\bar{\Gamma}\bar{Z}$ (C2 and C3 bands) that are nearly degenerate near $E_F$, while the C1 band is nearly dispersionless around -650 meV, suggesting its 1D nature. Along $\bar{\Gamma}\bar{M}$, the C3 band becomes nearly dispersionless, while the other two bands disperse strongly (Supplementary Fig. S4). The MDCs of the band structure along $\bar{\Gamma}\bar{Z}$ in Fig. 2d highlight the hybridization-induced band splitting between the C2 and C3 bands. These results suggest that the low-energy electronic structure is built by coupled 1D chains. The measured electronic structures are perfectly reproduced by the surface-projected *ab-initio* calculations of both the FS (Fig. 2e) and band dispersions (Fig. 2g), where the effect of weak inter-chain coupling is also well captured by the anti-crossings between the calculated FS sheets (Fig. 2f) and band splittings (Fig. 2h) (Supplementary Figs. S3 and S4).

We notice that the band dispersions, although cross $E_F$, exhibit strongly suppressed spectral weight near $E_F$ (Fig. 2c), which is reminiscent of the power-law spectral function of LL taking the hidden-1D electronic structures into account[38]. In addition, the bands disperse almost linearly near $E_F$, fulfilling the prerequisite for a LL phase (Supplementary Fig. S5). Figure 3a compares the integrated energy distribution curves (EDCs) of η-$Mo_4O_{11}$ with prototypical Fermi liquid and LL materials. The spectrum is clearly distinct from that of a Fermi-liquid system (e.g. the polycrystalline gold) with a Fermi edge but similar to that of the prototypical quasi-1D LL material $Li_{0.9}Mo_6O_{17}$[27] (Fig. 3a).

For a quantitative description of the LL behavior, we fit the experimental data to the finite-temperature LL model:

$$\rho(\epsilon, T) \propto T^\alpha \text{Re}\left[(2i)^{\alpha+1} B\left(\frac{\alpha+1+i\epsilon/\pi}{2}, -\alpha\right)\right], \quad (1)$$

where $\epsilon = (E - E_F)/k_B T$ is the scaled energy, B is the beta function, and $\alpha$ represents the LL anomalous exponent[39]. The analyses suggest $\alpha \approx 0$ for gold as expected for a Fermi liquid, and $\alpha \approx 0.69$ for $Li_{0.9}Mo_6O_{17}$, the same as that in a previous report[27] (Fig. 3a). Regarding η-$Mo_4O_{11}$, the ARPES spectra in a wide temperature range fit perfectly to the LL model with a temperature-independent $α ≈ 0.4 ± 0.1$ (Fig. 3b and Supplementary Fig. S6). In addition, as indicated by equation (1), the LL nature can be independently evidenced by the scaling behavior with temperature. That is, the ARPES intensity scaled by $T^α$ is temperature-independent, which is established excellently in Fig. 3c with $α ≈ 0.36$ (Supplementary Fig. S7). Moreover, the LL model suggests a linear dependence of $\log[I(E_F)]$ on $\log(T)$ with the slope being the LL exponent $α$ as plotted in Fig. 3d, from which we extract $α ≈ 0.38$, consistent with that obtained from the model fitting and scaling plot. Our findings thus compellingly prove the LL nature of η-$Mo_4O_{11}$ residing in its quasi-2D crystal structure.

To reveal the anisotropy of the unexpected LL behavior in the quasi-2D material, we measure ARPES spectra along different momentum directions, as indicated by the colored dashed lines in Fig. 3e. The integrated EDCs along different directions exhibit the same line shape and conform exactly to the LL model in all the measurements (Fig. 3f). Figure 3g,h summarize the $α$ values extracted independently from the model fitting and scaling plot, respectively. Within the experimental error, $α$ is independent of temperature and measurement direction, suggesting an in-plane isotropic LL behavior in the system.

To understand the novel LL behavior established above, we further investigate the electronic structure of η-$Mo_4O_{11}$ by combining *ab-initio* calculation and Wannier orbital construction in Fig. 4. The orbital-projected calculation demonstrates that the low-energy bands consist almost exclusively of 4$d$ orbitals of inner Mo atoms in the $Mo_6O_{22}$ slabs, i.e., Mo III and Mo IV in Fig. 1c (Supplementary Figs. S8 and S9). Specifically, the C1 and C2/C3 bands consist mainly of the Mo $d_{xy}$ and $d_{xz}$/$d_{yz}$ orbitals, respectively (Fig. 4a). For further exploration of the nature of these bands, a low-energy effective model based on Wannier orbitals is constructed, where the six bands that cross $E_F$ can be accurately described by the tight-binding model with six Wannier orbitals in a unit cell (Fig. 4a). The resulting Wannier orbitals form pairs, lying on either of the two slabs in a unit cell. The three pairs of Wannier orbitals feature different orientations and fall into two groups, as depicted from different viewing angles in Fig. 4b,c (Fig. 4d,e). All the orbitals are molecular-like and can be regarded as $d_{xy}$ and $d_{xz}/d_{yz}$ orbitals of Mo atoms. The Wannier orbital with $d_{xy}$ component fades out at Mo II and almost disappears at Mo I (the Mo tetrahedron layer), in agreement with the *ab-initio* calculations projected onto inequivalent Mo atoms (Supplementary Fig. S9) and thus builds a 1D chain along $b$, namely the C1 chain. The Wannier orbitals with $d_{xz}$ and $d_{yz}$ components also form chain-like structures extending along the $b+c$ and $b-c$ directions, which correspond to C2 and C3 chains and are related to each other by symmetry. In short, the Wannier orbitals feature identical orbital characters to those in *ab-initio* results (Fig. 4a) and extend along exactly the directions demanded within the hidden-1D scenario[40].

The observed hidden-1D electronic structure provides a natural explanation for many interesting properties of the quasi-2D η-$Mo_4O_{11}$, such as the periodic distortion in the lattice structure[41,42], FS nesting[36,43], and similarity with quasi-1D materials in optical spectra[44]. However, the robust LL behavior revealed in the current work is beyond the scope of canonical LL theory restricted to (quasi-)1D systems. Given the three arrays of chain-like structures in η-$Mo_4O_{11}$ that can host the 1D LL physics separately, the crossed-chain geometry naturally materializes the theoretical proposal of crossed LL phase (Fig. 1b), where a 2D LL phase exists in the phase space, especially at finite temperatures and for low-energy excitations[12,45].

It is noteworthy that the original proposal of crossed sliding LL is established in crossed chain arrays that are spatially separated[13]. In η-$Mo_4O_{11}$, however, all hidden-1D chains are constructed by the same octahedrons in the $Mo_6O_{22}$ slab (Fig. 4b-e) and thus cross through each other. To understand why the LL phase survives in such a 2D system, we conduct detailed analysis of the hopping amplitudes in the Wannier-orbital based tight-binding Hamiltonian, which reveals minor inter-chain coupling. Within the parallel chain array, the intra-chain nearest-neighbor hopping amplitude is 327 (232) meV for $d_{xy}$ ($d_{xz}/d_{yz}$) Wannier orbitals along the $b$ ($b+c/b-c$) direction, compared to the inter-parallel-chain and inter-crossed-chain hopping amplitude of only several tens meV (Supplementary Fig. S10 and Supplementary Table S1). We note that the hopping amplitudes within the two slabs in a unit cell differ from each other by several meV due to the inter-slab coupling, leading to the slight band splitting of the three band pairs in the *ab-initio* calculation (Fig. 4a), which is absent in a previous calculation based on the extended Hückel method[40] using the single-slab geometry. We emphasize that the 1D properties of the crossed chains, despite their overlapping in the real space, are preserved due to their orthogonal orbital components, which ingeniously realizes the intriguing crossed LL phase in η-$Mo_4O_{11}$ as schematically depicted in Fig. 4f (Supplementary Fig. S11).

The discovery of a crossed LL hidden in a quasi-2D material η-$Mo_4O_{11}$, not only materializes the long sought-after phase, but also sheds light on the understanding of non-Fermi liquid behaviors in other 2D and 3D materials.

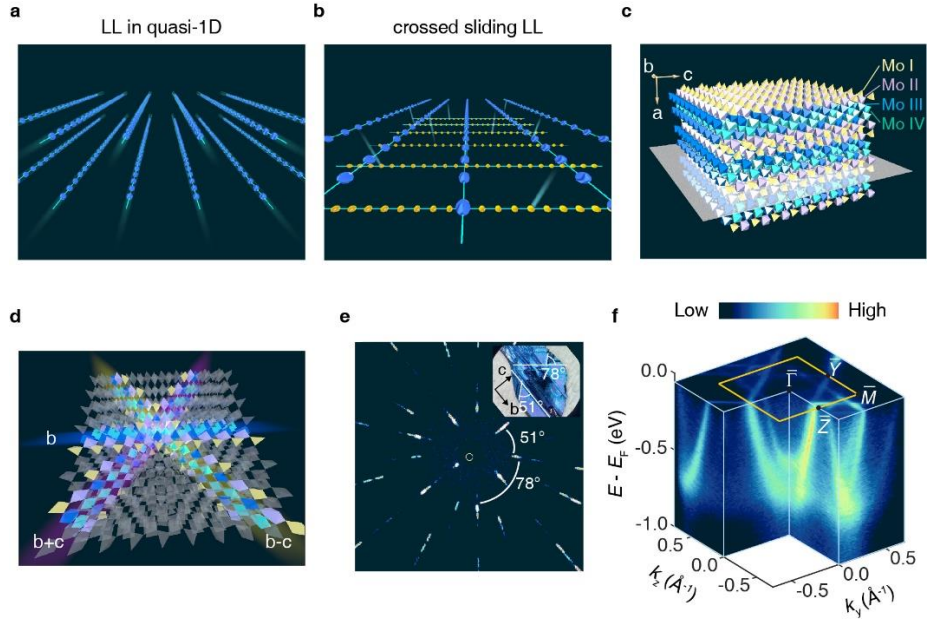

**Fig.1 | Physics in arrays of one-dimensional (1D) chains and basic properties of η-Mo$_4$O$_{11}$. a,** Schematic of a quasi-1D crystal consisting of parallel chains. **b,** Schematic of crossed arrays of planar parallel chains, which may host crossed sliding Luttinger liquid (LL) even with inter-array interaction. **c,** Overall crystal structure of η-Mo$_4$O$_{11}$, showing slabs consisted of MoO$_6$ octahedrons separated by MoO$_4$ tetrahedrons. The light gray plane indicates the (100) cleavage plane. Four inequivalent Mo atoms are marked. **d,** Projected view of a single slab in **c.** Selected chains along three different directions (*b*, *b+c*, *b-c*) are highlighted. (Also see Supplementary Fig. S1.) **e,** Laue diffraction pattern along the (100) plane. The inset shows a photo of a typical sample with three natural crystalline directions corresponding to the *b*, *b+c*, and *b-c* directions. **f,** Three-dimensional plot of the band structure measured at 30 K, using 50 eV photons. The yellow rectangle is the surface Brillouin zone (BZ).

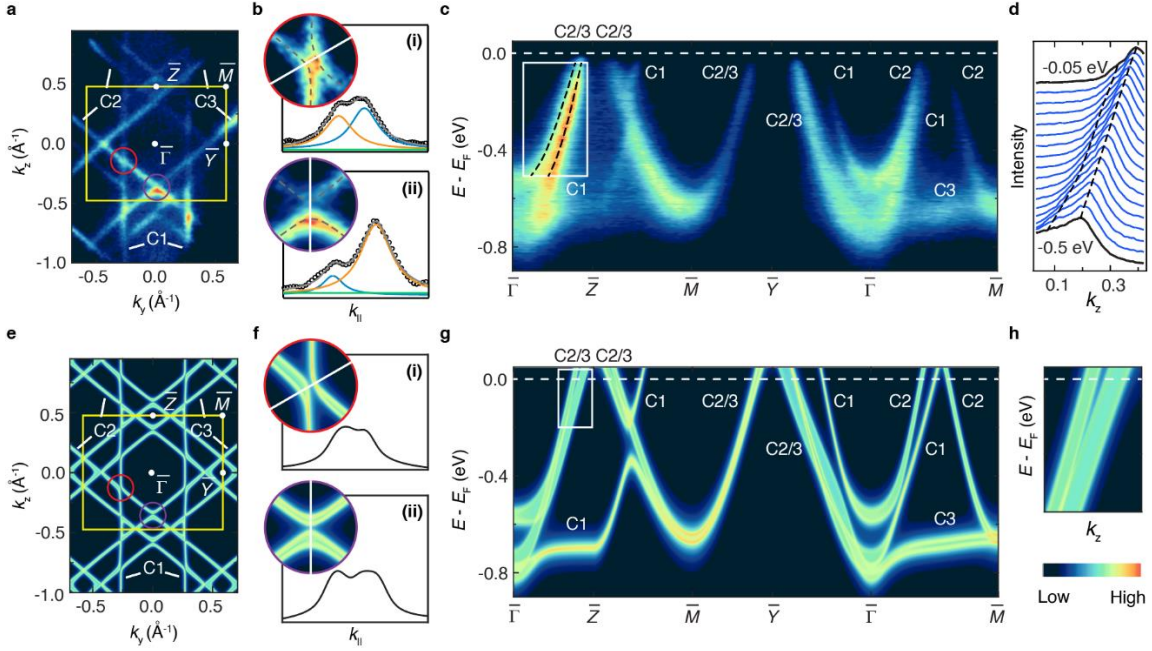

**Fig.2 | Comparison between the experimental and calculated electronic structure of η-Mo$_4$O$_{11}$.**
**a,** Measured Fermi surface (FS) showing three sets of quasi-1D sheets along different directions. The yellow rectangle indicates the first surface BZ. **b,** Magnification of the crossings between different FS sheets as indicated by the red and purple circles in **a**. The dashed gray lines are guides to eyes for FS sheets. The momentum distribution curves (MDC, black circles) along the white lines show the avoided crossings of FS sheets. The MDCs fit well to two Lorentzians (blue and red curves) on top of constant backgrounds (green curves). **c,** Experimental band structure along $\bar{\Gamma}\bar{Z}\bar{M}\bar{Y}\bar{\Gamma}\bar{M}$. The dashed white line indicates $E_F$. The dashed black lines are guides to eyes for the non-degenerate bands along $\bar{\Gamma}\bar{Z}$ induced by inter-chain coupling. **d,** Stacking plot of MDCs in the white rectangle in **c**. The dashed black lines are guides to eyes for the non-degenerate bands along $\bar{\Gamma}\bar{Z}$. Data in **a, c, d** were measured at 30 K with 50 eV photons, while data in **b** were measured at 15 K with 50 eV photons. **e-g,** Surface-projected *ab-initio* calculation of the electronic structure for comparing with **a-c**. The MDCs in **f** were convolved with experimental resolution. The white rectangle in **g** is enlarged in **h** for the non-degenerate bands along $\bar{\Gamma}\bar{Z}$. The calculated band structure is slightly renormalized by 1.1 to account for the minor correlation effect.

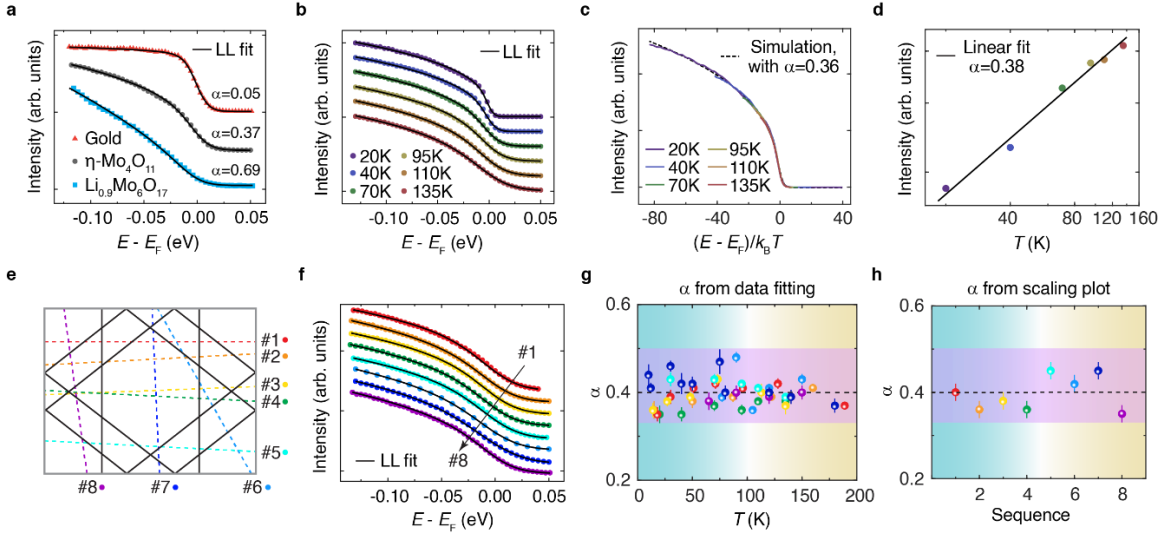

**Fig.3 | Luttinger liquid (LL) behavior in η-Mo$_4$O$_{11}$. a,** Comparison of integrated energy distribution curves (EDCs) of polycrystalline gold, Li$_{0.9}$Mo$_6$O$_{17}$, and η-Mo$_4$O$_{11}$. **b,** EDCs of η-Mo$_4$O$_{11}$ measured at different temperatures and the fits to the LL model. **c,** Scaling plot of the EDCs in **b** with a scaling factor of $T^{0.36}$. The dashed line plots the simulated LL spectral function with anomalous exponent $α = 0.36$. **d,** Log-log plot of the EDC intensity at $E_F$ versus temperature. The black line is a linear fit to the data with $α = 0.38$. **e,** ARPES measurement directions (dashed colored lines) and schematic FS sheets (solid black lines) in the first surface BZ (gray rectangle). **f,** EDCs measured along different directions as marked in **e** and the fits to the LL model. **g,** Value of $α$ extracted from data fitting at various temperatures in the measurements along different directions. **h,** Value of $α$ extracted from scaling plot in the measurements along different directions. In **g** and **h**, the colors of circles correspond to those in **e**, the dashed lines are the guides to eyes for the temperature independence of $α$, and the purple area highlights the scattering of $α$. Data in **a-d** were measured with 21.2 eV photons. Data in **f-h** were measured with photon energy of 21.2 eV (green, cyan, purple), 32 eV (red), 33 eV (orange, yellow), 50 eV (dark blue) and 55 eV (light blue).

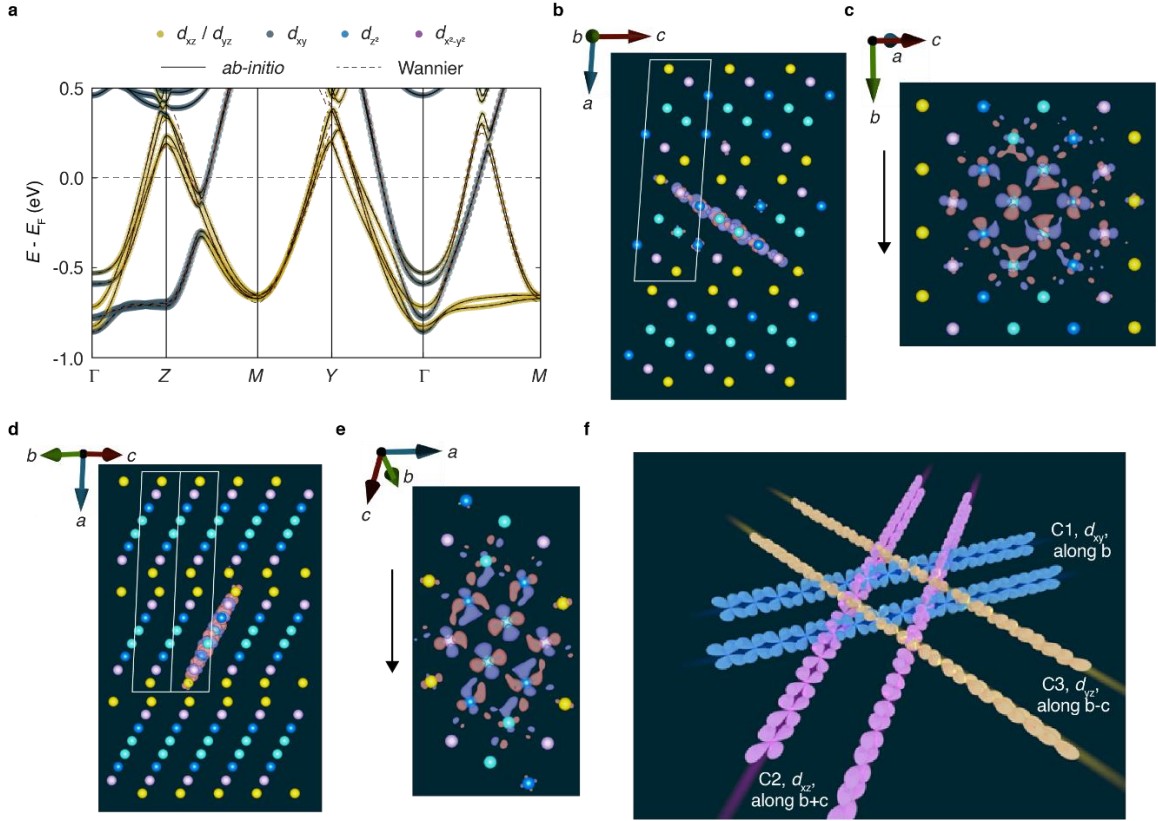

**Fig.4 | *Ab-initio* calculation of the electronic structure and Wannier orbitals of η-Mo$_4$O$_{11}$. a,** Comparison between band structure along high-symmetry directions obtained using *ab-initio* calculation (colored circles) and Wannier-orbital based tight-binding calculation (dashed brown lines). The size of colored circles is proportional to the weight of the corresponding orbital components. **c,d,** Viewing of Wannier orbital composed of $d_{xy}$ component along its extending (*b*) direction (**b**) and normal direction (**c**). **d,e,** Viewing of Wannier orbital composed of $d_{xz}$ component along its extending (*b+c*) direction (**d**) and normal direction (**e**). In **b-e**, for the Wannier orbitals, purple and red lobes denote positive and negative phases, respectively, and only Mo atoms are shown for clarity. The white lines in **b** and **d** indicate the unit cell. The black arrows in **c** and **e** indicate the extending direction of the quasi-1D Wannier orbitals. **f,** Schematic of the crossed LL phase preserved by the orthogonal orbitals of Mo 4*d* t$_{2g}$ orbitals.

**Methods**

**Sample growth and characterization.** Single crystals of η-$Mo_4O_{11}$ were grown by a chemical vapor transport technique as reported in previous work[46]. High-purity reagents $MoO_2$ (Alfa Aesar, 99.0%) and $MoO_3$ (Alfa Aesar, 99.998%) powder mixed with a molar ratio of about 1:3 were used as the starting materials. $TeCl_4$ (Alfa Aesar, 99.9%) was used as the transport agent. The mixtures were thoroughly ground and sealed into a quartz tube under a vacuum of $10^{-5}$ mbar, and subsequently placed in a two-zone furnace with a temperature profile of 560~510 °C to grow crystals. After maintaining this condition for 15 days, millimeter-sized η-$Mo_4O_{11}$ single crystals were obtained. For the X-ray diffraction (XRD) slice cuts, a Bruker D8 single-crystal X-ray diffractometer (SXRD) with Mo Kα ($\lambda$ = 0.71073 Å) was adopted at 150 K. The back-reflection Laue detector (MWL120, Multiwire Laboratories, Ltd.) was also used to check the quality of the obtained single crystals and determine the crystallographic orientation at room temperature.

**Angle-resolved photoemission spectroscopy.** High-resolution ARPES measurements were performed at beam line I05 of Diamond Light Source (DLS), beam line APE of Elettra Synchrotron Radiation Source, beam line 13U of National Synchrotron Radiation Laboratory (NSRL), beam line 03U of Shanghai Synchrotron Radiation Facility (SSRF), beam line 5-2 of Stanford Synchrotron Radiation Lightsource (SSRL) and Tsinghua University. Data were collected with Scienta R4000 electron analyzers at DLS, NSRL, SSRF, Scienta SES2002 electron analyzer at APE, and Scienta DA30L electron analyzer at SSRL and Tsinghua University. The overall energy and angle resolutions were set to 15 meV and 0.2°, respectively. The samples were cleaved *in situ* and measured under ultra-high vacuum below $1.0 \times 10^{-10}$ mbar.

*Ab-initio* **calculation.** *Ab-initio* band structure calculation was performed using QUANTUM ESPRESSO code package under Perdew-Burke-Ernzerhof (PBE) type generalized gradient approximation (GGA)[47,48]. Experimental structural parameters were relaxed with a force threshold of 0.01 eV/Å. The cutoff energy for the plane-wave basis was set to 600 eV and a Γ-centered Monkhorst-Pack *k*-point mesh of 11×14×3 was used to obtain a self-consistent charge density.

Fermi surface and (projected) density of states were calculated with a denser mesh of 22×28×6. The Wannier orbitals were constructed by the tight-binding based calculation supplied by the Wannier90 code[49,50], based on which the surface-projected and tight-binding band structures were calculated with WannierTools package[51].

**Acknowledgements**

This work is funded by the National Natural Science Foundation of China (Grants No. 11774190), the National Key R&D program of China (Grants No. 2017YFA0304600), and EPSRC Platform Grant (Grant No. EP/M020517/1).

**Competing interests**

Authors declare that they have no competing interests.

**Data and materials availability**

The datasets that support the findings of this study are available from the corresponding author on request.